\documentclass[12pt]{article}
\usepackage{amssymb,euscript,epsfig,color}
\newtheorem{lemma}{Lemma}
\newtheorem{theorem}[lemma]{Theorem}
\newtheorem{corollary}[lemma]{Corollary}

\newenvironment{proof}{{\bf Proof.}}{\hfill\rule{2mm}{2mm}}
\newenvironment{definition}{\noindent{\bf Definition.}}{}
\newtheorem{remarka}[lemma]{Remark}

\newcommand{\comment}[1]{ }
\newcommand{\peyman}[1]{#1}

\newcommand{\R}{\mathbb{R}}

\def\diam {{\rm diam}}
\begin{document}

\title{{\Large \bf Approximation and Inapproximability Results for Maximum Clique of Disc Graphs in High Dimensions}}

\author{
{\bf Peyman  Afshani$^a$ and Hamed Hatami$^b$} \\
{\small\it Department of Computer Science}\\
$^a${\small {University of Waterloo}}\\
$^b${\small University of Toronto}}
\date{}
\maketitle

\begin{abstract}
We prove algorithmic and hardness results for the problem of finding
the largest set of a fixed diameter in the Euclidean space. In
particular, we prove that if $A^*$ is the largest subset of diameter
$r$ of $n$ points in the Euclidean space, then for every
$\epsilon>0$ there exists a polynomial time algorithm that outputs a
set $B$ of size at least $|A^*|$ and of diameter at most
$r(\sqrt{2}+\epsilon)$. On the hardness side, roughly speaking, we
show that unless $P=NP$ for every $\epsilon>0$ it is not possible to
guarantee the diameter $r(\sqrt{4/3}-\epsilon)$ for $B$ even if the
algorithm is allowed to output a set of size $({95\over
94}-\epsilon)^{-1}|A^*|$.

\end{abstract}
\noindent {{\sc Keywords:} Approximation Algorithms; Computational
Geometry; Disc Graphs; Diameter.}

\thispagestyle{empty}
\setcounter{page}{1}

\section{Introduction}

The problem that we consider in this paper can be formulated as a
clustering problem. These types of problems have been studied for
quite long time and have many theoretical and practical applications
in computer science \cite{datta}. A branch of clustering problems
includes problems in which given a set of points the goal is to find
a ``cluster'' (or clusters) with minimum size or maximum number of
points. Typical examples of clusters include spheres, boxes, or any
other shape of fixed complexity. Of course, the difficulty of the
problem greatly depends on the definition of cluster. The clusters
that we consider here are all the shapes of constant diameter.
 The
\emph{diameter} of a set $S$ is defined as
$$\diam(S)= \sup_{x,y \in S} |x-y|,$$
where $|x-y|$ is the Euclidean distance between the two points $x$
and $y$. Specifically, we consider the following problem:
\begin{itemize}
\item[] {\bf Problem 1:} Let $P$ be a set of $n$ points in
$\mathbb{R}^d$ and $r>0$ be a real number. Find a subset $S
\subseteq P$ of maximum size which satisfies $\diam(S) \le r$.
\end{itemize}

A \emph{clique} is a graph in which every two vertices are adjacent.
For a graph $G$, let $\omega(G)$ denote the size of the maximum
clique in $G$, i.e. $\omega(G)$ is the maximum number of vertices of
$G$ such that every two of them are adjacent. Determining
$\omega(G)$ is called the \emph{maximum clique problem}. A closely
related topic is the notion of an independent set.  An
\emph{independent set} in $G$ is a subset of vertices such that no
two of them are adjacent. Let $\alpha(G)$ denote the size of the
maximum independent set in $G$. Determining $\alpha(G)$ is called
the \emph{maximum independent set problem}. Denote by $G^c$ the
complement of $G$, and note that $\omega(G)=\alpha(G^c)$. Thus, the
maximum clique problem for $G$ is equivalent to the maximum
independent set problem for $G^c$. For more on these topics we refer
the reader to~\cite{west}.

\comment{ Problem~1 is equivalent to the maximum clique problem in
\emph{disc graphs}: Build a graph $G$ with the vertex set $V(G)=P$
and connect two points $x,y$ if and only if $|x-y| \le r$. There is
a one to one correspondence between cliques in $G$ and sets of
diameter at most $r$ in $P$.
}

\peyman{Problem~1 is equivalent to the maximum clique problem in
\emph{disc graphs} which are defined as follows. Given a point set
$P \subset \mathbb{R}^d$ and a parameter $r$, a disc graph $G$ is
defined by $V(G)=P$ and $xy \in E(G)$ if and only if $|x-y| \le r$.
There is a one to one correspondence between cliques in $G$ and sets
of diameter at most $r$ in $P$. }

In Problem~1 the diameter is fixed and our objective is to maximize the
number of points. On the other hand, we can fix the number of points
and ask for the minimum diameter:

\begin{itemize}
\item[] {\bf Problem 2:} Let $P$ be a set of $n$ points in
$\mathbb{R}^d$ and $k>0$ be an integer. Find a subset $S \subseteq P$
of size $k$ with minimum diameter.
\end{itemize}

We show that both these problems are NP-hard when the dimension is
sufficiently large, i.e., for some $d = \Theta(\log n)$. In fact, we
prove a stronger result which shows that even certain approximations
of these problems are impossible unless P=NP. These approximations
are defined in the following way:

\begin{definition}
For $t,s \ge 1$, a $(t,s)$-approximation algorithm for Problem~1 is an algorithm
that returns a set $S$ of size at least $\mathit{Opt}/t$ so
that $\diam(S) \le sr$, where $\mathit{Opt}$ is the size of the
optimal answer to Problem~1.
\end{definition}

\begin{definition}
For $t,s \ge 1$, a $(t,s)$-approximation algorithm for Problem~2 is an algorithm
that returns a set $S$ of size at least $k/t$ so that $\diam(S) \le
s \times \mathit{Opt}$, where $\mathit{Opt}$ is the optimal answer
to Problem~2.
\end{definition}

These two problems are obtained by relaxing the size and diameter
constraints of the output set. A simple observation shows that
these two problems are equivalent.

\begin{lemma}
For $t,s \ge 1$, there exists a polynomial time $(t,s)$-approximation algorithm for
Problem 1 if and only if there exists a polynomial time $(t,s)$-approximation
algorithm for Problem 2.
\end{lemma}
\begin{proof}
Let ${\cal A}$ be a $(t,s)$-approximation algorithm for Problem 1.
Consider an instance $(P,k)$ of Problem 2. For every pair of
points $x,y \in P$, run ${\cal A}$ with parameter
$r_{x,y}:=|x-y|$. We output the minimum $r_{x,y}$ for which the
answer of ${\cal A}$ is of size at least $k/t$. Let $S_o$ be the
optimal solution to the $(P,k)$ instance of Problem 2. At some
point ${\cal A}$ is called with parameter $r:=\diam(S_o)$. Now the
output of ${\cal A}$ is a set of size at least $|S_o| / t \ge
k/t$ and with diameter at most $rs= s \times \diam(S_o)$.

To prove the other direction let ${\cal B}$ be a
$(t,s)$-approximation algorithm for Problem 2. Consider an
instance $(P,r)$ for Problem 1. A $(t,s)$-approximation algorithm
for Problem 1 can be obtained in a similar way by running ${\cal
B}$ for every $k=1,\ldots,n$.
\end{proof}

Since these two problems are equivalent we refer to both of them
as the Diameter Approximation Problem.

Both Problems 1 and 2 are solvable in the 2-dimensional plane in
polynomial time \cite{aggarwal,datta,polytopes}. For Problem~2 the
fastest known algorithm achieves the running time $O(n\log n +
k^{2.5}n\log k)$~\cite{polytopes}. It is shown in
\cite{cccg} that in the 3-dimensional space there is a $({\pi \over
\cos^{-1}1/3 },1)$-approximation algorithm. Finally, when the
dimension $d$ is a fixed constant, one can design a polynomial time
approximation scheme achieving a $(1,1+\epsilon)$-approximation, for
every $\epsilon>0$~\cite{cccg}. It is also easy to see that there
exists a trivial $(1,2)$-approximation algorithm for this problem in
any dimension: a ball with radius $r$ about a point $x \in P$
containing the maximum number of points is a $(1,2)$-approximation
for Problem 1. Thus, it is interesting to study at which point the
problem turns from polynomially solvable to NP-hard. We have the
following result in this direction:

\begin{theorem}
\label{thm:hardness} For every $\epsilon>0$ there exists
$d=\Theta(\log n)$, so that there is no polynomial time
$(\frac{95}{94}-\epsilon,\sqrt{4/3}-\epsilon)$-approximation
algorithm for the Diameter Approximation Problem in dimension $d$
unless P=NP.
\end{theorem}

We also improve upon the trivial $(1,2)$-approximation algorithm
and obtain the following theorem:

\begin{theorem}
\label{thm:approx} For every $\epsilon>0$ there is a polynomial time
$(1,\sqrt{2}+\epsilon)$-approximation algorithm for the Diameter
Approximation Problem in any dimension.
\end{theorem}

In Section~\ref{sec:hardness} we prove Theorem~\ref{thm:hardness}.
We use spectral properties to move from combinatorics of graphs to
geometry of Euclidean space. This technique in combination with a
hardness result regarding the maximum independent set problem in
$3$-regular graphs proves Theorem~\ref{thm:hardness}. In
Section~\ref{sec:approx} we prove Theorem~\ref{thm:approx} using
simple geometric techniques. Finally, in the ``Corollaries''
subsection we observe that having Theorem~\ref{thm:approx} in hand,
it is possible to move in the other direction.
Corollaries~\ref{cor:SDP} and~\ref{cor:eigenvalue} show that one can
apply Theorem~\ref{thm:approx} to the geometric representation of
the graph to approximate the maximum independent set problem for
certain graphs.

\section{From Graphs to Euclidean Space}

In this section we prove Theorem~\ref{thm:hardness}.
We show that unless P=NP, there is no
$(\frac{95}{94}-\epsilon,\sqrt{4/3}-\epsilon)$-approximation algorithm for the Diameter
Approximation Problem for every $\epsilon>0$.
We use spectral techniques to show that a certain metric on
the graph embeds isometrically into the Euclidean space.
This type of reduction from geometry to graph theory via metric
embedding has been successfully applied to various problems dealing
with the so called $(1-2)$-metrics~\cite{indyk,trevisan}.
Our proof of Theorem~\ref{thm:hardness} relies on the
following result:

\begin{theorem}{ \cite{3MIS}}
\label{thm:3reg} For every $\epsilon>0$, unless P=NP, there is no polynomial time algorithm
that approximates the maximum
independent set problem in $3$-regular graphs within a factor of $\frac{95}{94}-\epsilon$.
\end{theorem}

\subsection{Proof of Theorem~\ref{thm:hardness}}\label{sec:hardness}
To prove the theorem we
reduce the Diameter Approximation Problem to the maximum independent set problem in $3$-regular
graphs.

Consider a $3$-regular graph $G$. Denote by $G^c$ the complement of
$G$, and notice that cliques in $G^c$ correspond to independent sets
in $G$. We begin by finding a lower bound for the minimum eigenvalue
of the adjacency matrix $A_{G^c}$ of $G^c$. Denote by $\lambda_1 \le
\ldots \le \lambda_n$ the eigenvalues of $A_{G^c}$. Since $G^c$ is
an $(n-4)$-regular graph, its Laplacian can be written as
$L_{G^c}=(n-4)I - A_{G^c}$. Thus for a vector $v$, we have
$A_{G^c}v=\lambda v$ if and only if $L_{G^c}v=(n-4-\lambda)v$. This
shows that the eigenvalues of $L_{G^c}$ are $n-4-\lambda_n \le
\ldots \le n-4-\lambda_1$. It is well-known \cite{godsil} that the
maximum eigenvalue of the Laplacian is bounded by $n$. So the
minimum eigenvalue of $A_G$ is at least $-4$. Define
$Q:=A_{G^c}+(4+\gamma)I$ for $\gamma > 0$ chosen sufficiently small.
By the same argument that we used above to find the eigenvalues of
$L_{G^c}$, one can show that the smallest eigenvalue of $Q$ is
$\lambda_1 + 4+\gamma>0$ which means that $Q$ is a symmetric
positive definite matrix. This implies that there exists a
nonsingular matrix $U$, which can be found using elementary
techniques, such that $Q=U^tU$ (see~\cite{lalgebra} page 285 for
instance). Define a function $f:[n] \rightarrow \mathbb{R}^n$ by
setting the value of $f(i)$ to be the $i$-th row of the matrix $U$.
Note that
$$|f(i)|^2=f(i)\cdot f(i)=Q_{i,i}=4+\gamma,$$ and
$$|f(i)-f(j)|^2=|f(i)|^2+|f(j)|^2-2f(i)\cdot f(j)=8+2\gamma-2Q_{i,j}.$$
Thus
$$|f(i)-f(j)|=\left\{\begin{array}{ll}
    \sqrt{6+2\gamma} & A^c_{(ij)} = 1 \\
    \sqrt{8+2\gamma} &    A^c_{(ij)} = 0
\end{array}\right.$$
Consider the vertex $v_i$ of $V(G^c)$ which corresponds to the
$i$-th row and column of $A^c$.
Map $v_i$ to the point $f(i)$ in Euclidean space $\mathbb{R}^n$.
Let $P$ be the resulting point set in $\R^n$.
The above properties imply that every vertex of $V(G^c)$ is mapped to
a vector of magnitude $2$ and the distance between two vertices is $\sqrt{6+2\gamma}$, if
there is an edge between them, and $\sqrt{8+2\gamma}$ if not.


Using the Johnson-Lindenstrauss dimension reduction
lemma~\cite{dimreduction2,dimreduction} there exists a dimension
$d=O(\lambda^{-2}\log n)$, and a polynomial algorithm which maps $P$
into $\R^d$ such that the distance between any two points of $P$
changes by a factor of at most $1+\lambda/2$.
Let $g: V(G^c) \rightarrow \R^d$ be the corresponding map. Thus if
we choose $\lambda$ and $\gamma$ small enough, for every two
vertices $v_i, v_j \in V(G^c)$ the distance between $g(i)$ and
$g(j)$ is at most $(1+\lambda)\sqrt{6}$ if they are connected in
$G^c$ and at least $(1+\lambda)^{-1}\sqrt{8}$ if they are not
connected in $G^c$. So for a set $S \subset V(G^c)$, its geometric
representation will have diameter at most $(1+\lambda)\sqrt{6}$ if
it is a clique but it will have diameter at least
$(1+\lambda)^{-1}\sqrt{8}$ otherwise. By picking $\epsilon$ to be
small enough and applying a
$(\frac{95}{94}-\epsilon,\sqrt{4/3}-\epsilon)$-approximation
algorithm to Problem 1 with $g(G^c)$ and $r=\sqrt{6}$ we can find a
clique of size at least $1/(\frac{95}{94}-\epsilon)$ times the size
of the maximum clique in $G^c$. Theorem~\ref{thm:3reg} shows that
this is impossible unless $P=NP$.

\subsection{Proof of Theorem~\ref{thm:approx}}\label{sec:approx}
In this section we prove Theorem~\ref{thm:approx} by applying
simple geometric techniques. We follow the general ideas and
techniques of \cite{cccg,aggarwal}, borrowing and generalizing the
main tool from \cite{aggarwal}. The idea is to extend and
generalize the trivial $(1,2)$-approximation. One way to interpret
the $(1,2)$-approximation is to say that any set $A$ of diameter $r$
can be placed inside a ball of diameter $2r$ centered at a point in $A$. To obtain a
$(1,\sqrt{2}+\epsilon)$-approximation, we first show that any set of
diameter $r$ can actually be placed inside a ball of diameter
$(\sqrt{2}+\epsilon)r$, and then we produce a polynomial time algorithm to
compute such a ball.

Let $A$ be the optimal answer to Problem~1. We start by proving
that $A$ is inside a ball of diameter $(\sqrt{2}+\epsilon)r$. Let
$B(P,t)$ denote a ball of radius $t$ centered at point $P$.
At the beginning, $P_1$ is an arbitrary point of $A$ and thus
we have $A \subset B(P_1,r)$. At the $i$-th step we assume $A
\subset B(V_i,r_i)$ for a $V_i \in \mathbb{R}^d$ and some value
$r_i \le r$ to be determined later.

Let $P_{i+1}$ be the point of $A$ with maximum distance to $V_i$. This
implies,
$$A \subset B(V_i,r_i) \cap B(P_{i+1},r) \cap B(V_i,|V_i-P_{i+1}|)$$
and since $|V_i-P_{i+1}| \le r_i$, we have:
$$A \subset  B(P_{i+1},r) \cap B(V_i,|V_i-P_{i+1}|).$$

If $x=|V_i-P_{i+1}| \le r\sqrt{2}/2$ then the set of points is
inside a ball of diameter $\sqrt{2}r$. So we assume $x >
r\sqrt{2}/2$.

\begin{figure}
\begin{center}
    \ \input{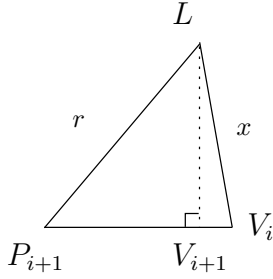}
\end{center}
\caption{The position  of $V_{i+1}$ with respect to the positions of $V_i$, $P_{i+1}$, and $L$. }
    \label{fig:step}
\end{figure}

Consider a point $L$ on the intersection of boundaries of the two
balls $B(P_{i+1},r)$ and $B(V_i,x)$ (Figure \ref{fig:step}).
Consider the plane passing through $L$, $P_{i+1}$ and $V_i$ and
draw the line $LV_{i+1}$ perpendicular to the segment
$P_{i+1}V_i$. A simple calculation proves that:
$$|L-V_{i+1}| = r \sqrt{1-{r^2\over 4x^2}}\le r \sqrt{1-{r^2\over 4r_i^2}}$$

Define $r_{i+1} = r \sqrt{1-{r^2\over 4r_i^2}}$. It can be also
verified that if $x > r \sqrt{2}/2$ then $|P_{i+1}-V_{i+1}| <
|V_{i+1}-L|$ and the ball $B(V_{i+1},|L-V_{i+1}|)$ will contain the
intersection $B(P_{i+1},r) \cap B(V_i,x)$. This implies $A \subset
B(V_{i+1},|L-V_{i+1}|) \subset B(V_{i+1},r_{i+1})$. It is easy to
check that the sequence $r_1, r_2, \cdots$ converges to
$r\sqrt{2}/2$. Thus given any $\epsilon>0$ it is possible to fix a
constant $k$ (depending only on $\epsilon$) such that $r_k < r\sqrt{2}/2 + \epsilon$.

To obtain an algorithm from the discussion above, we only need
to consider all different possible choices for $P_1,\ldots,P_k$.
Discarding the invalid choices or choices that result in an
invalid state, each choice for $P_1,\ldots,P_k$ leads to a ball
with radius at most $r\sqrt{2}/2+\epsilon$.
Now the algorithm outputs the one
which contains the maximum number of points. Since $k$ is a
constant, the algorithm is polynomial.

\subsection{Corollaries}
The following corollaries can be obtained using techniques employed
in the previous section.

\begin{corollary}
\label{cor:SDP} Fix $\delta>0$ and let $G$ be a graph such that
there exists a mapping $f:V(G)\rightarrow \mathbb{R}^n$ satisfying
$|f(u)-f(v)|
>\sqrt{2}+\delta $ if $uv \in E(G)$, and $|f(u)-f(v)| \le 1$ otherwise. Then
it is possible to find the size of the maximum independent set of
$G$ in polynomial time.
\end{corollary}
\begin{proof}
Let $V(G)=\{v_1,\ldots,v_n\}$.
Suppose that the $i$-th row of a matrix $M$ be the vector $f(v_i)$. Let further $A=MM^t$. Clearly,
$A$ is positive semi-definite, and $|f(v_i)-f(v_j)|^2=A_{ii}+A_{jj}-2A_{ij}$. Thus finding the map $f$
reduces to finding a positive semi-definite matrix $A$ with
$$A_{ii}+A_{jj}-2A_{ij} > (\sqrt{2}+\delta)^2 \qquad \qquad \forall v_iv_j \in E(G),$$
and
$$A_{ii}+A_{jj}-2A_{ij} \le 1 \qquad \qquad \forall i\neq j, v_iv_j \not\in E(G).$$
As in~(\cite{LLR95}, Theorem 3.2) the ellipsoid algorithm can be invoked to find such a matrix $A$.

Then we apply the algorithm of
Theorem~\ref{thm:approx} to $f(V(G))$ with the setting $r=1$ and $\epsilon=\delta/2$. Let $I$ be an independent
set of maximum size in $G$. Then the diameter of $f(I)$ is at most $1$ because $|f(u)-f(v)| \le 1$,
if $uv \not\in E(G)$. The algorithm given in the proof of Theorem~\ref{thm:approx}
finds a set $P$ of diameter $\sqrt{2}+\delta/2$ whose size is at least $|I|$. Since $|f(u)-f(v)|
>\sqrt{2}+\delta $ when $uv \in E(G)$, we conclude that $f^{-1}(P)$ is an independent set.
This completes the proof.
\end{proof}

\begin{corollary}
\label{cor:eigenvalue} Fix $\epsilon>0$ and let $G$ be a graph
whose minimum eigenvalue is at least $-2+\epsilon$. It is possible
to find the size of the maximum independent set of $G$ in
polynomial time.
\end{corollary}
\begin{proof}
By the proof of Theorem~\ref{thm:hardness} every such graph satisfies the condition of
Corollary~\ref{cor:SDP}.
\end{proof}

\section{Concluding Remarks}
\begin{itemize}
    \item We believe that Theorem~\ref{thm:approx} is ``almost''
    sharp in the sense that the constant $\sqrt{4/3}-\epsilon$ in
    Theorem~\ref{thm:hardness} can be improved to $\sqrt{2}-\epsilon$ becoming arbitrarily close to the $\sqrt{2}+\epsilon$
    upper bound of Theorem~\ref{thm:approx}.

    \item The main idea behind the proof of Theorem~\ref{thm:approx} was to introduce
    a polynomial time algorithm that given $n$ points
    computes a ball of diameter
    $(\sqrt{2}+\epsilon)r$ which contains the largest subset of the points that has diameter at most $r$.
    The fact that such a ball exists was already known, and in fact
    stronger results have been obtained using Helly-type arguments (we refer the reader to~\cite{DGK63,MatBook}
    for the proofs and the description of the Helly-type theorems). However, the novel part of the proof was
    the algorithmic aspect, and showing that there exists a polynomial time algorithm which  finds
    such a ball.

    \item There is already a body of work dedicated to characterization of
    all graphs with the smallest eigenvalue of at least
    $-2$ (see~\cite{bussemaker,lambda-2}). These graphs have been
    characterized as ``generalized line graphs'' plus some finite set
    of exceptions. This characterization gives an alternative
    proof for Corollary~\ref{cor:eigenvalue} which uses a different polynomial time algorithm.
\end{itemize}

\section*{Acknowledgements}
The authors would like to thank the anonymous referee for his/her
helpful comments and suggestions that greatly helped us to remove
some inaccuracies from the earlier version of this article.

\bibliographystyle{abbrv}
\bibliography{diam3}

\begin{thebibliography}{10}

\bibitem{cccg}
P.~Afshani and T.~Chan.
\newblock Approximation algorithms for maximum cliques in 3{D} {U}nit-{D}isk
  graphs.
\newblock In {\em Proceedings of the 17th Canadian Conference on Computational
  Geometry (CCCG'05)}, pages 19--22, 2005.

\bibitem{aggarwal}
A.~Aggarwal, H.~Imai, N.~Katoh, and S.~Suri.
\newblock Finding {$k$} points with minimum diameter and related problems.
\newblock {\em J. Algorithms}, 12(1):38--56, 1991.

\bibitem{bussemaker}
F.~C. Bussemaker and A.~Neumaier.
\newblock Exceptional graphs with smallest eigenvalue {$-2$} and related
  problems.
\newblock {\em Math. Comp.}, 59(200):583--608, 1992.

\bibitem{lambda-2}
P.~J. Cameron, J.-M. Goethals, J.~J. Seidel, and E.~E. Shult.
\newblock Line graphs, root systems, and elliptic geometry.
\newblock {\em J. Algebra}, 43(1):305--327, 1976.

\bibitem{3MIS}
J.~Chleb\'{\i}kov{\'a} and M.~Chleb\'{\i}k.
\newblock Inapproximability results for bounded variants of optimization
  problems.
\newblock {\em Electronic Colloquium on Computational Complexity (ECCC)},
  10(026), 2003.

\bibitem{DGK63}
L.~Danzer, B.~Gr{\"u}nbaum, and V.~Klee.
\newblock Helly's theorem and its relatives.
\newblock In {\em Proc. Sympos. Pure Math., Vol. VII}, pages 101--180. Amer.
  Math. Soc., Providence, R.I., 1963.

\bibitem{datta}
A.~Datta, H.-P. Lenhof, C.~Schwarz, and M.~Smid.
\newblock Static and dynamic algorithms for {$k$}-point clustering problems.
\newblock {\em J. Algorithms}, 19(3):474--503, 1995.

\bibitem{polytopes}
D.~Eppstein and J.~Erickson.
\newblock Iterated nearest neighbors and finding minimal polytopes.
\newblock {\em Discrete Comput. Geom.}, 11(3):321--350, 1994.

\bibitem{godsil}
C.~Godsil and G.~Royle.
\newblock {\em Algebraic graph theory}, volume 207 of {\em Graduate Texts in
  Mathematics}.
\newblock Springer-Verlag, New York, 2001.

\bibitem{indyk}
V.~Guruswami and P.~Indyk.
\newblock Embeddings and non-approximability of geometric problems.
\newblock In {\em SODA '03: Proceedings of the Fourteenth Annual ACM-SIAM
  Symposium on Discrete Algorithms}, pages 537--538, Philadelphia, PA, USA,
  2003. Society for Industrial and Applied Mathematics.

\bibitem{lalgebra}
F.~E. Hohn.
\newblock {\em Introduction to Linear Algebra}.
\newblock Macmillan, New York, 1972.

\bibitem{dimreduction2}
P.~Indyk and R.~Motwani.
\newblock Approximate nearest neighbors: towards removing the curse of
  dimensionality.
\newblock In {\em STOC '98: Proceedings of the Thirtieth Annual ACM Symposium
  on Theory of Computing}, pages 604--613, New York, NY, USA, 1998. ACM Press.

\bibitem{dimreduction}
W.~B. Johnson and J.~Lindenstrauss.
\newblock Extensions of {L}ipschitz mappings into a {H}ilbert space.
\newblock In {\em Conference in Modern Analysis and Probability (New Haven,
  Conn., 1982)}, volume~26 of {\em Contemp. Math.}, pages 189--206. Amer. Math.
  Soc., Providence, RI, 1984.

\bibitem{LLR95}
N.~Linial, E.~London, and Y.~Rabinovich.
\newblock The geometry of graphs and some of its algorithmic applications.
\newblock {\em Combinatorica}, 15(2):215--245, 1995.

\bibitem{MatBook}
J.~Matou{\v{s}}ek.
\newblock {\em Lectures on Discrete Geometry}, volume 212 of {\em Graduate
  Texts in Mathematics}.
\newblock Springer-Verlag, New York, 2002.

\bibitem{trevisan}
L.~Trevisan.
\newblock When {H}amming meets {E}uclid: the approximability of geometric {TSP}
  and {MST} (extended abstract).
\newblock In {\em STOC '97: Proceedings of the Twenty-Ninth Annual ACM
  Symposium on Theory of Computing}, pages 21--29, New York, NY, USA, 1997. ACM
  Press.

\bibitem{west}
D.~B. West.
\newblock {\em Introduction to graph theory}.
\newblock Prentice Hall Inc., Upper Saddle River, NJ, 1996.

\end{thebibliography}
\end{document}